# The current density and transport coefficients in the fully ionized plasma with *q*-distributions in nonextensive statistics

Wang Yue and Du Jiulin

*Department of Physics, School of Science, Tianjin University, Tianjin 300072, China*

**Abstract** We study the current density and transport coefficients in the fully ionized plasma with the *q*-distributions in nonextensive statistics and in strong magnetic field. By using the generalized Boltzmann transport equation in nonextensive statistics, we derive the current density and the expressions of the transport coefficients, including the conductivity, the thermoelectric coefficient, the Hall coefficient, and the Nernst coefficient. It is shown that these new transport coefficients has been generalized to the nonequilibrium complex plasmas with *q*-distributions in nonextensive statistics, which depend strongly on the *q*-parameters and when we take the limit *q*→1, they perfectly return to those for the plasma based on a Maxwellian distribution.

## I. Introduction

The power-law distributions are a class of non-Maxwellian distributions or non-exponential distributions that have been observed commonly in the fields of physics, astronomy, chemistry, life sciences and engineering technology, typical examples such as the *q*-distributions in complex systems described within the framework of nonextensive statistics [1], the *κ*-distributions observed in astrophysical and space plasmas and the *α*-distributions like the form of $E^{-\alpha}$ [2-8]. Theoretically, the power-law distributions can be found from a stochastic dynamics of the Brown's motion having a generalized fluctuation-dissipation relation (FDR) between the friction coefficient and the diffusion coefficient [9,10]. In plasmas physics, non-Maxwellian distributions have been observed and studied commonly both in astrophysical and space plasmas and in laboratory plasmas. For instance, the plasmas in the planetary magnetospheres are nonequilibrium and found to be far deviation from the Maxwellian distribution due to the presence of high energy particles [11]. The spacecraft measurements of plasma velocity distributions, both in the solar wind and in the planetary magnetospheres and magnetosheaths, have revealed that non-Maxwellian distributions are quite common. In many situations the velocity distributions have a "suprathermal" power-law tail at high energies, which has been well modeled by the famous *κ*-distribution [2,12]. In recent years, the researches on the complex plasmas with power-law *q*-distributions in nonextensive statistics have attracted great interest because their many interesting applications are found in the wide fields of space plasma physics and astrophysics [13-25]. In particular, the power-law distributions observed in plasmas can be studied under the framework of nonextensive statistics. And with the aid of the nonextensive *q*-kinetic theory, one can determine the expressions of the nonextensive *q*-parameter in the astrophysical and space plasmas and thereby understand its physical meaning [13,21-22].

Nonextensive statistics is a new statistical theory to generalize the traditional Boltzmann-Gibbs statistics that can be reasonably used to study complex systems, including the systems with long-range interactions, such as plasma and self-gravitating systems. On the basis of the nonextensive $q$-entropy and the probabilistically independent postulate, the nonextensive statistics is found to have the so-called $q$-distribution functions. Fore example, the velocity $q$-distribution function is described [1] by

$$f_q(\mathbf{v}) = nB_q \left(\frac{m}{2\pi k_B T}\right)^{\frac{3}{2}} \left[1 - (1-q)\frac{m(\mathbf{v}-\mathbf{u})^2}{2k_B T}\right]^{1/(1-q)},\tag{1}$$

where $q$ is a nonextensive parameter who's deviation from 1 represents the degree of nonextensivity, $T$ is temperature, $\mathbf{u}$ is the bulk velocity of the fluid under consideration, $n$ is the number density of particles, $m$ is mass of the particle, $k_B$ is Boltzmann constant, and $B_q$ is the $q$-dependent normalized constant given by

$$B_q = \begin{cases} (1-q)^{\frac{1}{2}}(3-q)(5-3q)\dfrac{\Gamma\left(\frac{1}{2}+\frac{1}{1-q}\right)}{4\Gamma\left(\frac{1}{1-q}\right)}, & \text{for } q \leq 1. \\[4mm] (q-1)^{\frac{3}{2}}\dfrac{\Gamma\left(\frac{1}{q-1}\right)}{\Gamma\left(\frac{1}{q-1}-\frac{3}{2}\right)}, & \text{for } q \geq 1. \end{cases}$$

In general, the temperature $T$, the density $n$ and the bulk velocity $\mathbf{u}$ in the complex plasma should be considered to be space inhomogeneous, i.e., $T=T(r)$, $n=n(r)$ and $\mathbf{u}=\mathbf{u}(r)$, and they can vary with time.

The transport coefficients in the power-law $\kappa$-distributed plasma were first studied under the Lorentz model [23], a simplified plasma model. Most recently, the diffusion and viscosity in the weakly ionized plasmas with the $\kappa$-distribution and the $q$-distribution were studied in [24] and [25], respectively. In the plasma, if the temperature and the density of a certain charged components in plasma are space inhomogeneous, there are electrons to move from a high temperature and high local density to a low temperature and low local density, leading the current. As important one of transport processes in plasma, the current describes the transport of electrons. In this work, we will study the current density and the transport coefficients in the fully ionized plasma with the velocity $q$-distributions in nonextensive statistics.

The paper is organized as follows. In Sec. 2, we introduce the generalized Boltzmann transport equations in nonextensive statistics. In Sec. 3, we study the current density in the $q$-distributed plasma and derive the transport coefficients in this physical situation in nonextensive statistics, including the Hall coefficient, the conductivity, the thermoelectric coefficient and the Nernst coefficient, respectively. Finally in Sec. 4, we give the conclusion.

## 2. The transport equations

The transport equation of the fully ionized plasma in strong magnetic field in nonextensive $q$-kinetic theory can be generalized by the Boltzmann equation,

$$\frac{\partial f_\alpha}{\partial t} + \mathbf{v} \cdot \nabla f_\alpha + \frac{Q_\alpha e}{m_\alpha} \left( \mathbf{E} + \frac{1}{c} \mathbf{v} \times \mathbf{B} \right) \cdot \nabla_v f_\alpha = C_{q,\alpha}(f_\alpha), \tag{2}$$

where $f_\alpha \equiv f_\alpha(\mathbf{r}, \mathbf{v}, t)$ is a single-particle velocity distribution function at time $t$, velocity $\mathbf{v}$ and position $\mathbf{r}$; $\nabla = \partial/\partial \mathbf{r}$; $\nabla_v = \partial/\partial \mathbf{v}$. $\mathbf{E}$ is the electric field, $\mathbf{B}$ is the magnetic field and $c$ is the light speed. On the right-hand side is the $q$-collision term, $C_{q,\alpha} = C_{q,\alpha\alpha} + C_{q,\alpha\beta}$, representing the change of the velocity distribution function due to the collisions between particles in the nonextensive $q$-kinetic theory. The subscript $\alpha = e$, $i$ and $\beta = e$, $i$ denote the electron and the ion respectively, and $Q_\alpha$ is the charge for $\alpha$th component in the plasma.

For the fully ionized plasma, $C_{q,\alpha}$ should be considered the Fokker-Planck (FP) $q$-collisions under the framework of nonextensive statistics. It has been proved [9] that in a complex system, if there is a generalized FDR between the diffusion coefficient $D_\alpha$ and the friction coefficient $\gamma_\alpha$, given for example by the condition,

$$D_\alpha = m_\alpha \gamma_\alpha k_B T_\alpha \left[ 1 - (1-q) \varepsilon_\alpha / k_B T_\alpha \right], \tag{3}$$

where $\varepsilon_\alpha$ is the energy, then the power-law $q$-distribution is the stationary-state solution of the FP equation, and the time-dependent solution of the FP equation will approach to the $q$-distribution in nonextensive statistics at long time [10]. Therefore, in the fully ionized plasma with the condition (3), the $q$-collision term $C_{q,\alpha}$ can be described by the FP collision operator. When we take the limit $q \to 1$, the condition (3) becomes the standard FDR and the solution of FP equation return to a Maxwell-Boltzmann distribution in the traditional statistics.

According to the Chapman-Enskog expansion [26], we can write the velocity distribution function of the $\alpha$th component as the following form,

$$f_\alpha = f_\alpha^{(0)} + f_\alpha^{(1)} + f_\alpha^{(2)} + \cdots, \tag{4}$$

where $f_\alpha^{(r)}$ is a small disturbance of the stationary-state distribution $f_\alpha^{(0)}$, and with $f_\alpha^{(r)} \ll f_\alpha^{(r-1)}$ for $r = 1, 2, \ldots$. In Eq.(4) for the present situation, the velocity $q$-distribution (1) can be the stationary state distribution $f_\alpha^{(0)}$. And therefore, $f_\alpha^{(0)}$ can be written as

$$f_\alpha^{(0)} = f_{q,\alpha} = n_\alpha B_q \left( \frac{m_\alpha}{2\pi k_B T_\alpha} \right)^{\frac{3}{2}} \left[ 1 - (1-q) \frac{m_\alpha (\mathbf{v} - \mathbf{u})^2}{2 k_B T_\alpha} \right]^{\frac{1}{1-q}}, \tag{5}$$

Because the current density only contains the contribution of electrons, the current density in the power-law $q$-distributed plasma is given by

$$\mathbf{J}_q = -e \int \mathbf{v} f_e(\mathbf{v}) d\mathbf{v} = -e \int \mathbf{v} \left[ f_{q,e}(\mathbf{v}) + f_{q,e}^{(1)}(\mathbf{v}) + f_{q,e}^{(2)}(\mathbf{v}) + \ldots \right] d\mathbf{v}. \tag{6}$$

And because $f_e^{(0)}(\mathbf{v}) = f_{q,e}(\mathbf{v})$ is an even function of $\mathbf{v}$, the integral for $f_{q,e}(\mathbf{v})$ vanishes in (6). And only if the small disturbance $f_{q,e}^{(r)}(\mathbf{v})$, $r = 1, 2, 3 \ldots$, is an odd function of $\mathbf{v}$, the integrals on the right side of (6) are not zero.

According to Eq.(2), the transport equation of electrons [27] becomes

$$\frac{\partial f_e}{\partial t} + \mathbf{v} \cdot \nabla f_e - \frac{e}{m_e}\left(\mathbf{E} + \frac{1}{c}\mathbf{v} \times \mathbf{B}\right) \cdot \nabla_v f_e = C_{q,ee} + C_{q,ei}, \tag{7}$$

where $C_{q,ee}$ and $C_{q,ei}$ are the electron-electron and electron-ion collisions, respectively, in the $q$-kinetic theory.

## 3. The current density and the transport coefficients

In the stationary state, the distribution function of electrons is $f_e^{(0)} = f_{q,e}$, so we have that

$$C_{q,ee}\left(f_e^{(0)}, f_e^{(0)}\right) = C_{q,ee}\left(f_{q,e}, f_{q,e}\right) = 0, \tag{8}$$

$$C_{q,ei}\left(f_e^{(0)}, f_i^{(0)}\right) = C_{q,ei}\left(f_{q,e}, f_{q,i}\right) = 0, \tag{9}$$

Substituting (4) into Eq.(7) and considering that $f_\alpha^{(r)} \ll f_\alpha^{(r-1)}$ and

$$\frac{e}{cm_e}\left(\mathbf{v} \times \mathbf{B}\right) \cdot \nabla_v f_{q,e} = 0, \tag{10}$$

we can write Eq.(7) as

$$\frac{\partial f_{q,e}}{\partial t} + \mathbf{v} \cdot \nabla f_{q,e} - \frac{e}{m_e}\mathbf{E} \cdot \nabla_v f_{q,e} = \frac{e}{cm_e}\left(\mathbf{v} \times \mathbf{B}\right) \cdot \nabla_v \left(f_{q,e}^{(1)} + f_{q,e}^{(2)} + \ldots\right) + I_q\left(f_{q,e}^{(1)} + f_{q,e}^{(2)} + \ldots\right), \tag{11}$$

where $I_q(\ldots)$ is the collision integral operator. Using the $q$-distribution (5) in the nonextensive kinetic theory, this equation further becomes

$$\begin{aligned}
&\frac{m_e}{k_B T_e}\left[1 - (1-q)\frac{m_e v^2}{2k_B T_e}\right]^{-1} f_{q,e}\mathbf{v} \cdot \frac{\partial \mathbf{u}}{\partial t} + \mathbf{v} \cdot \nabla f_{q,e} - \frac{e}{m_e}\mathbf{E} \cdot \nabla_v f_{q,e} \\
&= \frac{e}{m_e c}\left(\mathbf{v} \times \mathbf{B}\right) \cdot \nabla_v \left(f_{q,e}^{(1)} + f_{q,e}^{(2)} + \ldots\right) + I_q\left(f_{q,e}^{(1)} + f_{q,e}^{(2)} + \ldots\right).
\end{aligned} \tag{12}$$

Because the transport coefficients considered here have nothing to do with the bulk velocity $\mathbf{u}$, without loss of generality, we can consider the case of $\mathbf{u}=0$. So Eq.(12) can be simplified as

$$\mathbf{v} \cdot \nabla f_{q,e} - \frac{e}{m_e}\mathbf{E} \cdot \nabla_v f_{q,e} = \frac{e}{m_e c}\left(\mathbf{v} \times \mathbf{B}\right) \cdot \nabla_v \left(f_{q,e}^{(1)} + f_{q,e}^{(2)} + \ldots\right) + I_q\left(f_{q,e}^{(1)} + f_{q,e}^{(2)} + \ldots\right). \tag{13}$$

In the plasma, the general form of the Ohm's law can be expressed [27] by

$$\mathbf{E} + \frac{1}{c}\mathbf{u} \times \mathbf{B} + \frac{1}{en_e}\nabla p_e = \frac{\mathbf{J}_{//}}{\sigma_{//}} + \frac{\mathbf{J}_\perp}{\sigma_\perp} + \Re\mathbf{B} \times \mathbf{J} + \alpha_{//}\nabla_{//}T + \alpha_\perp\nabla_\perp T + \aleph\mathbf{B} \times \nabla T, \tag{14}$$

where $\mathbf{J}_{//}$ and $\sigma_{//}$ are the current density and the conductivity, respectively, that are parallel to the magnetic field; $\mathbf{J}_\perp$ and $\sigma_\perp$ are the current density and the conductivity, respectively, that are perpendicular to the magnetic field; $\alpha_{//}$ and $\alpha_\perp$ are the thermoelectric coefficient that are respectively parallel and perpendicular to the magnetic field; $p_e$ is the pressure of the electrons, $\Re$ is the Hall coefficient, and $\aleph$ is the Nernst coefficient.

Without loss of generality, we still let **u**=0 for the convenience of calculations. The Ohm's law (20) in nonextensive statistics is in parallel written as

$$\mathbf{E} + \frac{1}{en_e}\nabla p_{q,e} = \frac{\mathbf{J}_{q//}}{\sigma_{q//}} + \frac{\mathbf{J}_{q\perp}}{\sigma_{q\perp}} + \Re_q\mathbf{B}\times\mathbf{J}_q + \alpha_{q//}\nabla_{//}T + \alpha_{q\perp}\nabla_{\perp}T + \aleph_q\mathbf{B}\times\nabla T .\qquad(15)$$

Usually, because the transport parallel to the magnetic field direction is the same as that without magnetic field, we only consider the transverse transport perpendicular to the magnetic field direction, then Eq.(21) can be reduced to

$$\mathbf{E}_{\perp} + \frac{1}{en_e}\nabla_{\perp}p_{q,e} = \frac{\mathbf{J}_{q\perp}}{\sigma_{q\perp}} + \Re_{q\perp}\mathbf{B}\times\mathbf{J}_{q\perp} + \alpha_{q\perp}\nabla_{\perp}T + \aleph_{q\perp}\mathbf{B}\times\nabla_{\perp}T .\qquad(16)$$

### 3.1 *The first-order approximate equation for the current density*

Because the longitudinal transport coefficient is independent of $\mathbf{B}$, here we only discuss the transverse transport coefficient, perpendicular to the magnetic field direction. Introducing the Larmor frequency $\omega_{Be}$ of the electrons,

$$\omega_{Be} = \frac{eB}{m_e c},\qquad(17)$$

Eq. (13) is therefore written as

$$\mathbf{v}\cdot\nabla f_{q,e} - \frac{e}{m_e}\mathbf{E}\cdot\nabla_v f_{q,e} = \omega_{Be}\left(\mathbf{v}\times\hat{\mathbf{B}}\right)\cdot\nabla_v\left(f_{q,e}^{(1)}+f_{q,e}^{(2)}+\dots\right) + I_q\left(f_{q,e}^{(1)}+f_{q,e}^{(2)}+\dots\right),\quad(18)$$

where $\hat{\mathbf{B}} = \mathbf{B}/B$ is a unit vector along the direction of **B**. If the two terms on the right side of Eq.(18), for the magnetic field and collision, are considered of the same order, in the first-order approximation, this equation can be simplified [28] by

$$\mathbf{v}\cdot\nabla f_{q,e}^{(0)} - \frac{e}{m_e}\mathbf{E}\cdot\nabla_v f_{q,e}^{(0)} = \omega_{Be}\left(\mathbf{v}\times\hat{\mathbf{B}}\right)\cdot\nabla_v f_{q,e}^{(1)} + I_q\left(f_{q,e}^{(1)}\right),\qquad(19)$$

where the operator $I_q$ can only contain electron-ion collisions because electron-electron collisions have no contribution to the current. Generally, the collision operator can be approximated by the Lorentz operator [28], i.e., $I_q\left(f_{q,e}^{(1)}\right) = -\nu_{ei}(v)f_{q,e}^{(1)}$, with the electron-ion collision frequency given by

$$\nu_{ei}(v) = \frac{4\pi ze^4 n_e L_e}{m_e^2 v_e^3},\qquad(20)$$

and

$$L_e \approx \ln\left(\frac{\lambda_e m_e v_e^2}{ze^2}\right),$$

where $ze$ is the charge of the ion, $\lambda_e$ is Debye length of the electron and $v_e$ is the speed of the electron [27].

Usually, the velocity distribution function may assumed to be axisymmetric about the direction of the current, so that we can write

$$f_{q,e}^{(1)} = \frac{\mathbf{v}\cdot\mathbf{f}_{q,e}^{(1)}}{v}, \text{ and}\qquad(21)$$

$$\nabla_v f_{q,e}^{(1)} = \nabla_v \left( \frac{\mathbf{v} \cdot \mathbf{f}_{q,e}^{(1)}}{v} \right) = \frac{\mathbf{v}\mathbf{v}}{v^2} \cdot \frac{\partial \mathbf{f}_{q,e}^{(1)}}{\partial v} + \left( \frac{\mathbf{1}}{v} - \frac{\mathbf{v}\mathbf{v}}{v^3} \right) \cdot \mathbf{f}_{q,e}^{(1)} \quad . \tag{22}$$

Substituting into the right side of Eq.(19) we find that

$$\nabla f_{q,e}^{(0)} - \frac{e}{vm_e} \mathbf{E} \frac{\partial}{\partial v} f_{q,e}^{(0)} = \frac{\omega_{Be}}{v} \hat{\mathbf{B}} \times \mathbf{f}_{q,e}^{(1)} - \frac{v_{ei}}{v} \mathbf{f}_{q,e}^{(1)} \quad . \tag{23}$$

Here we only consider the transverse transport perpendicular to the magnetic field direction because the transport parallel to the magnetic field direction is the same as that without the magnetic field. If we write $\mathbf{f}_{q,e}^{(1)}$ as two parts, $\mathbf{f}_{q,e}^{(1)} = \mathbf{f}_{q,e\parallel}^{(1)} + \mathbf{f}_{q,e\perp}^{(1)}$, being the vectors parallel to and perpendicular to the magnetic field respectively, and then taking the vector product on the both sides of Eq.(23) by $\hat{\mathbf{B}}$, we have that in the transverse transport,

$$\hat{\mathbf{B}} \times \nabla_\perp f_{q,e}^{(0)} - \frac{e}{vm_e} \left( \hat{\mathbf{B}} \times \mathbf{E}_\perp \right) \frac{\partial}{\partial v} f_{q,e}^{(0)} = -\frac{\omega_{Be}}{v} \mathbf{f}_{q,e\perp}^{(1)} - \frac{v_{ei}}{v} \hat{\mathbf{B}} \times \mathbf{f}_{q,e\perp}^{(1)}. \tag{24}$$

Using Eq,(23) we can further write Eq.(24) as

$$\left( \omega_{Be}^2 + v_{ei}^2 \right) \frac{\mathbf{f}_{q,e\perp}^{(1)}}{v} = \omega_{Be} \left( -\hat{\mathbf{B}} \times \nabla_\perp f_{q,e}^{(0)} + \frac{e}{vm_e} \hat{\mathbf{B}} \times \mathbf{E}_\perp \frac{\partial}{\partial v} f_{q,e}^{(0)} \right)$$

$$-v_{ei} \left( \nabla_\perp f_{q,e}^{(0)} - \frac{e}{vm_e} \mathbf{E}_\perp \frac{\partial}{\partial v} f_{q,e}^{(0)} \right). \tag{25}$$

And then, from (21) we find that

$$f_{q,e}^{(1)} = \frac{\mathbf{v} \cdot \mathbf{f}_{q,e\perp}^{(1)}}{v} = \frac{\omega_{Be}}{\omega_{Be}^2 + v_{ei}^2} \left[ -\mathbf{v} \cdot \left( \hat{\mathbf{B}} \times \nabla_\perp f_{q,e}^{(0)} \right) + \frac{e}{m_e} \left( \hat{\mathbf{B}} \times \mathbf{E}_\perp \right) \cdot \nabla_{v,\perp} f_{q,e}^{(0)} \right]$$

$$-\frac{v_{ei}}{\omega_{Be}^2 + v_{ei}^2} \left( \mathbf{v} \cdot \nabla_\perp f_{q,e}^{(0)} - \frac{e}{m_e} \mathbf{E}_\perp \cdot \nabla_{v,\perp} f_{q,e}^{(0)} \right). \tag{26}$$

Because the gyrofrequency $\omega_{Be}$ is usually much larger than the collision frequency $v_{ei}$, this equation can be simplified as

$$f_{q,e}^{(1)} \approx \frac{1}{\omega_{Be}} \left[ \frac{e}{m_e} \left( \hat{\mathbf{B}} \times \mathbf{E}_\perp \right) \cdot \nabla_{v,\perp} f_{q,e}^{(0)} - \mathbf{v} \cdot \left( \hat{\mathbf{B}} \times \nabla_\perp f_{q,e}^{(0)} \right) \right]$$

$$-\frac{v_{ei}}{\omega_{Be}^2} \left( \mathbf{v} \cdot \nabla_\perp f_{q,e}^{(0)} - \frac{e}{m_e} \mathbf{E}_\perp \cdot \nabla_{v,\perp} f_{q,e}^{(0)} \right). \tag{27}$$

Therefore, according to Eq.(6), the first-order approximation for the current density can be obtained (see Appendix) by

$$\mathbf{J}_{q,e\perp}^{(1)} = -e \int \mathbf{v} f_{q,e}^{(1)}(\mathbf{v}) d\mathbf{v} = \frac{n_e e^2}{m_e \omega_{Be}} \hat{\mathbf{B}} \times \left[ \mathbf{E}_\perp + \frac{2k_B}{e(7-5q)} \nabla_\perp T_e \right]$$

$$+ \frac{\sqrt{2} e^2 n_e v_{ei}(T_e) B_q}{3\sqrt{\pi} m_e \omega_{Be}^2} \left[ \mathbf{E}_\perp - \frac{k_B}{2e(2-q)} \nabla_\perp T_e \right], \quad 0 < q < \frac{7}{5}, \tag{28}$$

where $\nu_{ei}(T_e)$ is the collision frequency at temperature $T_e$ in a Maxwellian velocity distribution, given by

$$\nu_{ei}(T_e) = \frac{4\pi z e^4 n_e L_e}{m_e^{1/2}(k_B T_e)^{3/2}}. \tag{29}$$

## 3.2 The transport coefficients

In the nonextensive statistics, the plasma gas pressure can be expressed [29], for the electrons, by

$$p_{q,e} = \frac{2}{(7-5q)} n_e k_B T_e, \quad 0 < q < \frac{7}{5}. \tag{30}$$

Thus, substituting the current density Eq(28) as well as the pressure Eq.(30) into Eq.(16), in the case that only the transverse transports, i.e., perpendicular to the magnetic field, are considered and in the first-order approximation, we find the following equations:

$$\mathbf{E}_\perp = \left[ \frac{1}{\sigma_{q\perp}} \frac{\sqrt{2\pi} e^2 n_e \nu_{ei}(T_e) B_q}{3\pi m_e \omega_{Be}^2} - ecn_e \Re_{q\perp} + O\left(\frac{1}{\omega_{Be}^2}\right) \right] \mathbf{E}_\perp, \tag{31}$$

$$\left[ \frac{1}{\sigma_{q\perp}} \frac{en_e c}{B} + \Re_{q,\perp} \frac{\sqrt{2} e^2 n_e \nu_{ei}(T_e) B B_q}{3\sqrt{\pi} m_e \omega_{Be}^2} + O\left(\frac{1}{\omega_{Be}^2}\right) \right] \hat{\mathbf{B}} \times \mathbf{E}_\perp = 0, \tag{32}$$

$$\left[ -\frac{2k_B}{e(7-5q)} - \frac{1}{\sigma_{q\perp}} \frac{k_B e n_e \nu_{ei}(T_e) B_q}{3\sqrt{2\pi}(2-q) m_e \omega_{Be}^2} - \Re_{q\perp} \frac{2cn_e k_B}{(7-5q)} + \alpha_{q\perp} + O\left(\frac{1}{\omega_{Be}^3}\right) \right] \nabla_\perp T_e = 0, \tag{33}$$

$$\left[ \frac{1}{\sigma_{q\perp}} \frac{2cn_e k_B}{(7-5q)B} - \Re_{q\perp} \frac{en_e \nu_{ei}(T_e) k_B B B_q}{3\sqrt{2\pi}(2-q) m_e \omega_{Be}^2} + \aleph_{q\perp} B + O\left(\frac{1}{\omega_{Be}^2}\right) \right] \hat{\mathbf{B}} \times \nabla_\perp T_e = 0, \tag{34}$$

From Eq.(31) and Eq.(32) we find the Hall coefficient,

$$\Re_{q\perp} = -\frac{1}{n_e ec \left(1 + \frac{2\nu_{ei}^2 B_q^2}{9\pi \omega_{Be}^2}\right)} + O\left(\frac{1}{\omega_{Be}^3}\right), \tag{35}$$

and the conductivity coefficient,

$$\sigma_{q\perp} = \frac{3\sqrt{\pi} e^2 n_e}{\sqrt{2} m_e \nu_{ei} B_q} \left(1 + \frac{2\nu_{ei}^2 B_q^2}{9\pi \omega_{Be}^2}\right) + O\left(\frac{1}{\omega_{Be}^3}\right). \tag{36}$$

And then substituting (35) and (36) into Eq. (33), we find the thermoelectric coefficient,

$$\alpha_{q\perp} = \frac{(5-3q)k_B \nu_{ei}^2 B_q^2}{3(7-5q)(2-q)\pi e \omega_{Be}^2} \left(1 + \frac{2\nu_{ei}^2 B_q^2}{9\pi \omega_{Be}^2}\right)^{-1} + O\left(\frac{1}{\omega_{Be}^3}\right), \tag{37}$$

and substituting (35) and (36) into Eq. (34), we find the Nernst coefficient,

$$\aleph_{q\perp} = -\frac{(5-3q)\nu_{ei}k_B B_q}{(7-5q)(2-q)\sqrt{2\pi}m_e c\omega_{Be}^2}\left(1+\frac{2\nu_{ei}^2 B_q^2}{9\pi\omega_{Be}^2}\right)^{-1} + O\left(\frac{1}{\omega_{Be}^3}\right). \tag{38}$$

In the above expressions (35)-(38), we have denoted $\nu_{ei} \equiv \nu_{ei}(T_e)$. It is clear that the transverse transport coefficients in the present plasma depend strongly on the $q$-parameter in nonextensive statistics, and when we take the limit $q\to 1$ they recover the standard expressions in the plasma with the Maxwellian distributions perfectly.

## 4. Conclusion and discussion

In conclusion, we have studied the current density and the transport coefficients in the fully ionized plasma with the power-law $q$-distributions in nonextensive statistics and with a strong magnetic field. Because the transport parallel to the magnetic field direction is the same as that without magnetic field, we only consider the transverse transport perpendicular to the magnetic field direction.

Based on the generalized Boltzmann transport equation in the $q$-kinetic theory and the power-law velocity $q$-distributions, where the $q$-collision terms are considered the Fokker-Planck collisions in the power-law distributions, the current density perpendicular to the magnetic field in the $q$-distributed plasma has been derived in the first-order approximation, given by Eq.(28).

Further, we applied the extended expression of the Ohm's law to the fully ionized plasma with the power-law $q$-distributions and thus derived the new expressions of the transverse transport coefficients from the current density in nonextensive statistics, including the Hall coefficient in Eq.(35), the conductivity coefficient in Eq.(36), the thermoelectric coefficient in Eq.(37) and the Nernst coefficient in Eq.(38). It is shown that these new transport coefficients all depend strongly on the $q$-parameter in nonextensive statistics, and if we take the limit $q\to 1$, they perfectly return to those in the case of the plasma with the Maxwellian velocity distribution.

On the nonextensivity, i.e., the degree of the $q$-parameter deviation from unity $q\neq 1$, in the nonequilibrium plasma with the $q$-distribution and with a magnetic field, its physical meaning has been understood by the relation [13, 21],

$$k_B\nabla T_e = (q-1)e\left(\nabla\varphi_c - c^{-1}\mathbf{u}\times\mathbf{B}\right), \tag{39}$$

where $\varphi_c$ is the Coulomb potential. Thus, the nonextensive $q$-parameter different from unity has represented correctly the properties of the nonequilibrium complex plasma with the Coulomb long-range interactions as well as electromagnetic interactions between the charged particles. These transport coefficients obtained in this paper can be applicable to study the fully ionized plasmas with the above nonequilibrium physical properties. Further information on the physical meaning of the $q$-parameter in nonequilibrium astrophysical and space plasmas may be found in [22].

Finally, we state that the statistical averages used in this paper are based on the standard definition of average because the four versions of Tsallis statistics has been proved to be equivalent and the standard average has been considered as good as any of the others [30]. Thus the $q$-dependent equations in this work and the derived $q$-dependent transport coefficients (35)-(38) have been generalized in nonextensive statistics, which are in parallel to the traditional results of kinetic theory and the fluid dynamics.

**Acnowledgement**

This work is supported by the National Natural Science Foundation of China under Grant No. 11775156.

**Appendix**

For $q > 1$, Eq. (28) is written as

$$\mathbf{J}_{q,e\perp}^{(1)} = -e\int \mathbf{v}_{\perp}f_{q,e\perp}^{(1)}d\mathbf{v}$$

$$= \frac{e}{\omega_{Be}}\int \mathbf{v}_{\perp}\left[\mathbf{v}\cdot\left(\mathbf{B}\times\nabla_{\perp}f_{q,e}^{(0)}\right) - \frac{e}{m_e}\left(\mathbf{B}\times\mathbf{E}_{\perp}\right)\cdot\nabla_{v,\perp}f_{q,e}^{(0)}\right]d\mathbf{v}$$

$$+ \frac{e}{\omega_{Be}^2}\int \mathbf{v}_{\perp}\nu_{ei}\left(\mathbf{v}\cdot\nabla_{\perp}f_{q,e}^{(0)} - \frac{e}{m_e}\mathbf{E}_{\perp}\cdot\nabla_{v,\perp}f_{q,e}^{(0)}\right)d\mathbf{v}$$

$$= \frac{e}{3\omega_{Be}}\left[\mathbf{B}\times\nabla_{\perp}\left(\int v^2 f_{q,e}^{(0)}d\mathbf{v}\right) - \frac{e}{m_e}\left(\mathbf{B}\times\mathbf{E}_{\perp}\right)\int v\frac{\partial}{\partial v}f_{q,e}^{(0)}d\mathbf{v}\right]$$

$$+ \frac{e}{3\omega_{Be}^2}\left[\nabla_{\perp}\left(\int \nu_{ei}v^2 f_{q,e}^{(0)}d\mathbf{v}\right) - \frac{e}{m_e}\mathbf{E}_{\perp}\int \nu_{ei}v\frac{\partial}{\partial v}f_{q,e}^{(0)}d\mathbf{v}\right]$$

$$= \frac{4\pi e n_e B_q}{3\omega_{Be}}\left\{\mathbf{B}\times\nabla_{\perp}\left(\left(\frac{m_e}{2\pi k_B T_e}\right)^{3/2}\int_0^{+\infty}v^4 A_q dv\right) + \left(\mathbf{B}\times\mathbf{E}_{\perp}\right)\frac{e}{k_B T_e}\left(\frac{m_e}{2\pi k_B T_e}\right)^{3/2}\int_0^{+\infty}v^4 A_q^{-1}dv\right\}$$

$$+ \frac{16\pi^2 z e^5 n_e L_e B_q}{3 m_e^2 \omega_{Be}^2}\left\{\nabla_{\perp}\left(\left(\frac{m_e}{2\pi k_B T_e}\right)^{3/2}\int_0^{+\infty}v A_q d\mathbf{v}\right) + \mathbf{E}_{\perp}\frac{e}{k_B T_e}\left(\frac{m_e}{2\pi k_B T_e}\right)^{3/2}\int_0^{+\infty}v A_q^{-1}d\mathbf{v}\right\},$$

where we have denoted $A_q \equiv \left[1 - (1-q)\frac{m_e v^2}{2 k_B T_e}\right]^{1/(1-q)}$. After the integrals in the above equation are completed, we have that

$$\mathbf{J}_{q,e\perp}^{(1)} = \frac{n_e e^2}{m_e \omega_{Be}}\hat{\mathbf{B}}\times\left[\mathbf{E}_{\perp} + \frac{2k_B}{(7-5q)e}\nabla_{\perp}T_e\right]$$

$$+ \frac{4\sqrt{2\pi}z e^6 n_e^2 L_e B_q}{3\omega_{Be}^2\left(m_e k_B T_e\right)^{3/2}}\left[\mathbf{E}_{\perp} - \frac{k_B}{2e(2-q)}\nabla_{\perp}T_e\right], \quad 1 \leq q < \frac{7}{5}.$$

In the same way, for $q < 1$, if $a = \sqrt{\frac{2k_B T_e}{m_e(1-q)}}$, then Eq. (28) can be written as

$$\mathbf{J}_{q,e\perp}^{(1)} = \frac{4\pi e n_e B_q}{3\omega_{Be}}\left\{\mathbf{B}\times\nabla_{\perp}\left(\left(\frac{m_e}{2\pi k_B T_e}\right)^{3/2}\int_0^a v^4 A_q dv\right) + \left(\mathbf{B}\times\mathbf{E}_{\perp}\right)\frac{e}{k_B T_e}\left(\frac{m_e}{2\pi k_B T_e}\right)^{3/2}\int_0^a v^4 A_q^{-1}dv\right\}$$

$$+ \frac{16\pi^2 z e^5 n_e L_e B_q}{3 m_e^2 \omega_{Be}^2}\left\{\nabla_{\perp}\left(\left(\frac{m_e}{2\pi k_B T_e}\right)^{3/2}\int_0^a v A_q d\mathbf{v}\right) + \mathbf{E}_{\perp}\frac{e}{k_B T_e}\left(\frac{m_e}{2\pi k_B T_e}\right)^{3/2}\int_0^a v A_q^{-1}d\mathbf{v}\right\}.$$

After the integrals in this equation are completed, we have that

$$\mathbf{J}_{q,e\perp}^{(1)} = \frac{n_e e^2}{m_e \omega_{Be}} \hat{\mathbf{B}} \times \left[ \mathbf{E}_\perp + \frac{2k_B}{(7-5q)e} \nabla_\perp T_e \right]$$

$$+ \frac{4\sqrt{2\pi} z e^6 n_e^2 L_e}{3\omega_{Be}^2 (m_e k_B T_e)^{3/2}} B_q \left[ \mathbf{E}_\perp - \frac{k_B}{2e(2-q)} \nabla_\perp T_e \right], \quad 0 < q \le 1.$$

## References

[1] C. Tsallis, Introduction to Nonextensive Statistical Mechanics: Approaching a Complex World, Springer, New York, 2009.

[2] S. R. Cranmer, Astrophys. J. **508** (1998) 925 and the references there in.

[3] R. G. DeVoe, Phys. Rev. Lett. **102** (2009) 063001.

[4] L. K. Gallos and P. Argyrakis, Phys. Rev. E **72** (2005) 017101.

[5] J. R. Claycomb, D. Nawarathna, V. Vajrala and J. H. Miller, J. Chem. Phys. **121** (2004) 12428.

[6] R. W. Benz, S. J. Swamidass and P. Baldi, J. Chem. Inf. Model **48** (2008) 1138.

[7] A. Hasegawa, K. Mima and M. Duong-van, Phys. Rev. Lett. **54** (1985) 2608.

[8] G. Boffetta, V. Carbone, P. Giuliani, P. Veltri and A. Vulpiani, Phys. Rev. Lett. **83** (1999) 662.

[9] J. Du, J. Stat. Mech. (2012) P02006.

[10] R. Guo and J. Du, Ann. Phys. **359** (2015)187.

[11] J. D. Mihalov, J. H. Wolfe, and. L. A. Frank, J. Geophys. Res. **81** (1976) 3412.

[12] V. M. Vasyliunas, J. Gerophys. Res. **73** (1968) 2839.

[13] J. Du, Phys. Lett. A **329** (2004) 262.

[14] M. P. Leubner, Astrophys. J. **604** (2004) 469.

[15] L. Y. Liu and J. L. Du, Physica A **387** (2008) 4821.

[16] Z. P. Liu, L. Y. Liu and J. L. Du, Phys. Plasmas **16** (2009) 072111.

[17] Z. E. Abbasi and A. Esfandyari-Kalejahi, Phys. Plasmas **23** (2016) 073112.

[18] M. Bacha, L. A. Gougam and M. Tribeche, Physica A **466** (2017) 199.

[19] A. Merriche and M. Tribeche, Ann. Phys. **376** (2017) 436.

[20] B. S. Chahal, M. Singh, Shalini and N.S. Saini, Physica A **491** (2018) 935.

[21] H. Yu and J. Du, Ann. Phys. **350** (2014) 302.

[22] H. Yu and J. Du, EPL **116** (2016) 60005.

[23] J. Du, Phys. Plasmas **20** (2013) 092901.

[24] L.Wang and J. Du, Phys. Plasmas **24** (2017) 102305; Phys. Plasmas **26** (2019) 019902.

[25] Y. Wang and J. Du, Phys. Plasmas **25** (2018) 062309.

[26] S. Chapman and T. G. Cowling, The Mathematical Theory of Non-uniform Gases: An Account of the Kinetic Theory of Viscosity, Thermal Conduction and Diffusion in Gases, Cambridge University Press, 1970.

[27] Z. Q. Huang and E. J. Ding, Transport Theory, Science Press, Beijing, 2008.

[28] P. Helander and D. J. Sigmar, Collisional Transport in Magnetized Plasmas, Cambridge University Press, 2002.

[29] J. Du, CEJP **3** (2005) 376.

[30] G. L. Ferri, S. Martinez and A. Plastino, J. Stat. Mech. (2005) P04009.